\documentclass{article}
\usepackage{amssymb}
\usepackage{amsmath}
\usepackage{graphicx}
\usepackage{float}

\begin{document}

\title{Dynamical Casimir effect and the possibility of laser-like generation
of gravitational radiation}
\date{December 22, 2017}
\author{R.Y. Chiao,$^{1}$ J.S. Sharping,$^{2}$ L.A. Martinez, B.S. Kang,
\and A. Castelli, N. Inan, and J.J. Thompson\medskip \\
%EndAName
Dept. of physics, University of California at Merced\\
$^{1}$rchiao@ucmerced.edu, $^{2}$jsharping@ucmerced.edu}
\maketitle

\textbf{Abstract:} In this paper, we address the question as to whether or
not measurable sources for gravitational waves could possibly be made in the
laboratory. Based on an analogy of the dynamical Casimir effect with the
stimulated emission of radiation in the laser, our answer to this question
is in the affirmative, provided that superconducting radio-frequency
cavities in fact possess high quality factors for both electromagnetic and
gravitational microwave radiation, as one would expect due to a
quantum-mechanical gravitational Meissner-like effect. In order to
characterize the response of matter to \textit{tensor} gravitational fields,
we introduce a prefactor to the source term of the gravitational wave
equation, which we call the \textquotedblleft relative gravitational
permeativity\textquotedblright\ analogous to the \textquotedblleft relative
electric permittivity\textquotedblright\ and \textquotedblleft relative
magnetic permeability\textquotedblright\ that characterize the \textit{vector%
} response of matter to applied fields in electromagnetism. This allows for
a possibly large quantum mechanical enhancement of the response of a superconductor
to an incident tensor gravitational wave field. Finally, we describe our experimental work
with high-Q superconducting radio-frequency cavities, and propose a design
for a coupled-cavity system with a flexible superconducting membrane in its
middle as its amplifying element. This will then allow us to test for a
Meissner-like expulsion, and therefore reflection, of incident tensor gravitational wave fields, and, above a
certain threshold, to generate coherent gravitational radiation via the
dynamical Casimir effect.

\medskip

\textbf{Text:} The 2017 Nobel prize in physics was awarded for observations
of gravitational waves arising from the inspiral of black hole pairs \cite%
{Nobel 2017}\cite{LIGO}. Recently, the emission of gravitational waves was
also observed due to the inspiral of a pair of neutron stars, along with the
simultaneous observations of gamma ray and optical detections of the same
event from the same source \cite{neutron star inspiral}.

The question naturally arises: Is it possible to generate gravitational
radiation in the laboratory? A common response to this question is the one
given by Misner, Thorne, and Wheeler (MTW), in their classic text \cite{MTW}:

\begin{quote}
\textquotedblleft The construction of a laboratory generator of
gravitational radiation is a non attractive enterprise in the absence of new
engineering or a new idea or both.\textquotedblright
\end{quote}

This response was a result of Einstein's calculation of the power emitted in
gravitational radiation $P_{\text{Einstein}}$ by a rotating steel beam,
which was based on his quadrupole formula \cite{MTW}\cite{LxL}%
\begin{equation}
P_{\text{Einstein}}=\frac{G}{45c^{5}}\left\langle \dddot{Q}%
_{ij}^{2}\right\rangle  \label{Einstein quadrupole formula}
\end{equation}%
where $G$ is Newton's constant, $c$ is the speed of light, and where \cite%
{LxL}%
\begin{equation}
Q_{ij}=\int \rho \left( 3x_{i}x_{j}-\delta _{ij}x^{k}x_{k}\right) dV
\label{mass quadrupole formula}
\end{equation}%
is the mass quadrupole tensor (Einstein's summation convention is being used
here, with Latin indices denoting spatial dimensions).

Einstein \cite{Einstein1918} calculated the gravitational radiation emitted
from a massive steel beam with a length of 20 meters and a radius of 1
meter, whose mass is 4.9$\times 10^{5}$ kilograms, rotating end-over-end
around its midpoint at its maximum possible angular velocity near its
breaking point, which is determined by the tensile strength of steel, 3$%
\times 10^{6}$ J m$^{-2}$. Then the maximum possible angular velocity of the
steel beam due to its tensile strength is 28 radians per second, and the
gravitational radiation power predicted by the quadrupole formula (\ref%
{Einstein quadrupole formula}) turns out to be%
\begin{equation}
P_{\text{Einstein}}\simeq 2.2\times 10^{-29}\text{ Watts}
\label{10e-(29) watts}
\end{equation}%
which is a miniscule amount of power. The basic reason for this arises from
the fact that the prefactor%
\begin{equation}
G/c^{5}\sim 10^{-53}\left( \text{Watts}\right) ^{-1}  \label{prefactor}
\end{equation}%
in Einstein's quadrupole formula (\ref{Einstein quadrupole formula}), is an
extremely small number. This is a consequence of the fact that Newton's
constant $G$, which is a tiny number to begin with, is combined with the
inverse $quintic$ power of the speed of light $c$, which is yet a much
tinier number.

Put differently, there arises a characteristic power $P_{\text{GR}}$ in
general relativity which is given by the fundamental constants $G$ and $c$
in the combination%
\begin{equation}
P_{\text{GR}}\equiv \frac{c^{5}}{G}=3.6\times 10^{52}\text{ Watts}
\end{equation}%
As pointed out by MTW in the beginning of their classic text \cite{MTW}, the
only place where such enormous powers could occur naturally is in
astrophysical sources, such as in supernova explosions. In fact, the first
direct observation of gravitational waves by LIGO \cite{LIGO}, was in the
merger of a pair of massive black holes orbiting each other at relativistic
speeds, an extreme case of an astrophysical source. Thus it would seem
impossible, for all practical purposes, for gravitational radiation power to
ever be produced in laboratory sources.

However, note that Planck's constant $\hbar $ is absent from Einstein's
quadrupole formula (\ref{Einstein quadrupole formula}) for the emission of
gravitational radiation. Could the \textquotedblleft new engineering or a
new idea or both,\textquotedblright\ as suggested in the above quotation
from MTW, be \textquotedblleft $quantum$ engineering,\textquotedblright\ in
which $\hbar $ somehow replaces $G$ and $c$, so that the necessity for the
use of astrophysical sources for the generation of gravitational waves could
somehow be avoided? Here we propose a possibly affirmative answer to this
question that involves the laser-like generation of gravitational radiation
via the process of the dynamical Casimir effect \cite{Aharonov paper}.

The starting point for this new \textquotedblleft quantum
engineering\textquotedblright\ approach to the generation of gravitational
waves is the assumption that the uncertainty principle%
\begin{equation}
\Delta E\Delta t\geq \frac{\hbar }{2}
\end{equation}%
leads to the existence of zero-point fluctuations with the zero-point energy%
\begin{equation}
E_{0}=\frac{1}{2}\hbar \omega  \label{zpe}
\end{equation}%
for $any$ kind of wave which oscillates with a frequency $\omega $. In
particular, this zero-point energy should apply to gravitational waves, as
well as to electromagnetic waves. In the case of gravitational waves, note
that the size of the zero-point energy (\ref{zpe}) is independent of
Newton's constant $G$ and of the speed of light $c$. Rather, it depends
solely on Planck's constant $\hbar $ and the frequency $\omega $. Although $%
\hbar $ is a tiny number, its tininess can be compensated for by the
exponential growth of the gravitational wave arising from the process of
stimulated emission of radiation, just like in the case of the laser.

Stimulated emission of gravitational-wave quanta, i.e., of gravitons,
follows from a quantum treatment of the radiation oscillators \cite{Garrison
& Chiao} that result from a $linearization$ of the theory of general
relativity \cite{Chiao1961}, in which the metric tensor $g_{\mu \nu }$ is
decomposed into the Minkowski metric tensor components $\eta _{\mu \nu }=$
diag $\left( -1,+1,+1,+1\right) $, which are large, and the metric deviation
tensor components $h_{\mu \nu }$, which are small, viz.,%
\begin{equation}
g_{\mu \nu }=\eta _{\mu \nu }+h_{\mu \nu }
\end{equation}%
The small, simple harmonic motion of the $linearized$ gravitational
radiation oscillators can be quantized by means of the standard quantization
condition%
\begin{equation}
\left[ a_{G},a_{G}^{\dag }\right] =1  \label{canonical commutator}
\end{equation}%
where $a_{G}$ is the annihilation operator for a given gravitational
radiation oscillator, and $a_{G}^{\dag }$ is the creation operator for the
same radiation oscillator (all other commutators for nonidentical radiation
oscillators being set equal to zero). It follows from the canonical
commutator (\ref{canonical commutator}) that%
\begin{equation}
a_{G}^{\dag }\left\vert n_{G}\right\rangle =\sqrt{n_{G}+1}\left\vert
n_{G}+1\right\rangle  \label{recursion relation}
\end{equation}%
where $\left\vert n_{G}\right\rangle $ is the number state containing $n_{G}$
gravitons in a given mode of the radiation field (i.e., an excitation of a
given radiation oscillator with $n_{G}$ quanta), and $\left\vert
n_{G}+1\right\rangle $ is the number state containing $n_{G}+1$ gravitons.
As Feynman has pointed out in \cite{Feynman Lectures}, the creation of an
extra radiation quantum with the probability amplitude of $\sqrt{n_{G}+1}$
in the recursion relationship (\ref{recursion relation}) leads to the
process of stimulated emission of radiation. Hence the recursion
relationship (\ref{recursion relation}) implies the possibility of the
laser-like generation of gravitational radiation.

\begin{figure}[]
\begin{center}
\includegraphics[width=0.8026\textwidth]{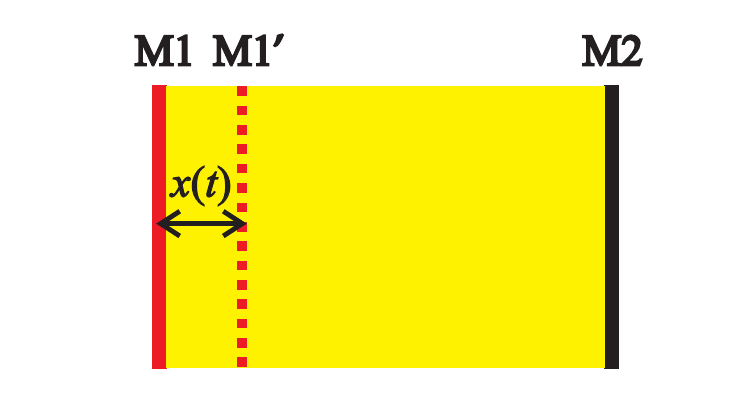} 
\end{center}
\caption{Sketch of the dynamical Casimir effect ("DCE"). A Fabry-Perot
cavity consists of two parallel mirrors M1 and M2. Mirror M1 is moving back
and forth sinusoidally with a time-dependent displacement $x(t)$ relative to
the mirror M2, which is a stationary mirror. The piston-like pumping action
of M1 upon the vacuum fluctuations contained inside the Fabry-Perot cavity
amplifies them parametrically so that they will become macroscopically
observable radiation (indicated in yellow) that fills up this resonator.}
\label{dynamicalcasimireffect}
\end{figure}

Figure 1 is an illustration of the dynamical Casimir effect
(\textquotedblleft DCE\textquotedblright ), in which a moving mirror M1
(red) of an initially empty Fabry-Perot interferometer moves sinusoidally
with a displacement $x\left( t\right) $ relative to a fixed mirror M2 \cite%
{Moore}\cite{Fulling&Davies}. The back-and-forth motion of mirror M1 is like
the back-and-forth motion of a piston that can do work on the vacuum
fluctuations which are contained within the two mirrors of the Fabry-Perot
cavity, thus amplifying them into laser-like light (yellow) via the process
of stimulated emission of radiation. However, unlike in an ordinary laser,
there is no need here for the introduction of a medium consisting of
two-level atoms with inverted populations in between the two mirrors of the
Fabry-Perot, since the push-and-pull mechanical pumping motions of the
mirror M1, in conjunction with the vacuum fluctuations in EM or GR radiation
fields as \textquotedblleft seed radiation,\textquotedblright\ are
sufficient\ for the laser-like, coherent generation of both EM and GR kinds
of radiations.

In other words, even if the Fabry-Perot resonator in Figure 1 were to be
initially totally empty except for vacuum fluctuations, the pumping action
of the mechanically moving mirror upon these fluctuations will cause the
resonator to fill up with radiation (indicated by the yellow region in
between M1 and M2 in Figure 1), seemingly \textquotedblleft out of
nothing,\textquotedblright\ just as coherent light is seemingly generated
\textquotedblleft out of nothing\textquotedblright\ in a laser above its
threshold. This is because the action of the moving mirror is like the
action of a moving piston which pushes and pulls on a gas of photons or
gravitons contained within the resonator. Thus the piston can impart energy
into this gas. As a result, the action of the piston can parametrically
amplify the radiation contained inside the resonator \cite{swing}, so that
it can become, via an exponential growth mechanism \cite{Aharonov paper}, a
macroscopically observable beam of coherent radiation, just like in a laser.

\begin{figure}[]
\begin{center}
\includegraphics[width=0.95\textwidth]{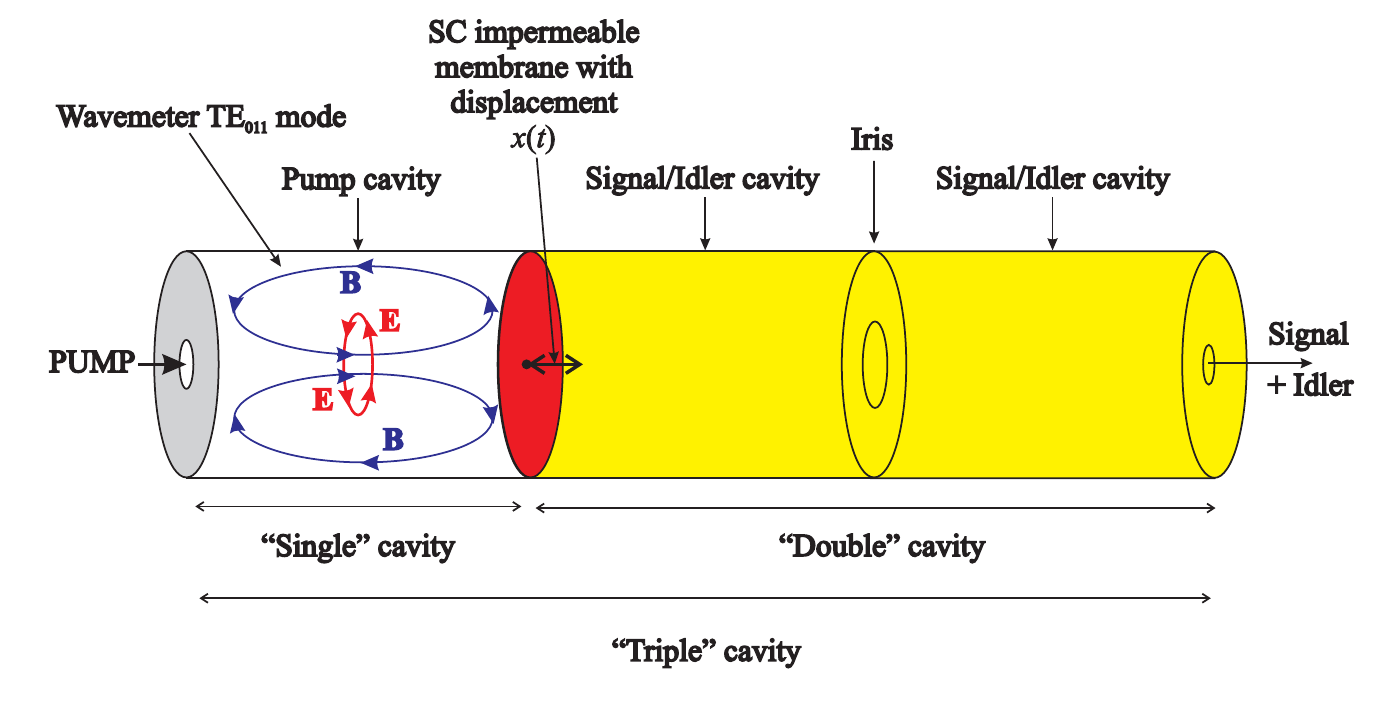} 
\end{center}
\caption{\textquotedblleft Triple\textquotedblright\ cavity parametric amplifier (paramp) is
divided into three adjacent superconducting (SC) cylindrical chambers. The
first chamber (i.e., the pump cavity) is separated from the second and third
chambers (i.e., the signal and idler cavities) by an impermeable SC membrane
(red), which is the active element of the paramp that can be driven into
motion with a displacement of $x(t)$ by radiation pressure from a pump wave
of the \textquotedblleft single\textquotedblright\ cavity. The initially
empty SC \textquotedblleft double cavity\textquotedblright\ on its right
side then fills up with radiation (yellow) that grows exponentially out of
vacuum fluctuations in the DCE.}
\label{triplecavity}
\end{figure}

Nation et al \cite{nation} have pointed out that the quantum amplification
process in the dynamical Casimir effect (DCE) is equivalent to the
amplification process in a parametric amplifier (paramp), such as the one in
the \textquotedblleft triple\textquotedblright\ cavity paramp configuration
illustrated in Figure 2, in which a membrane is pumped into mechanical
motion by the radiation pressure from pump microwaves in the leftmost
\textquotedblleft single\textquotedblright\ cavity. This membrane moves like
the moving mirror M1 in Figure 1 with a sinusoidal displacement $x\left(
t\right) $ that amplifies the signal and idler waves inside the
\textquotedblleft double\textquotedblright\ cavity on the right side of the
membrane \cite{paramp kinematics}.

For this and similar paramps, Nation et al \cite{nation} give the following
threshold:%
\begin{equation}
v_{\max }\geq \frac{c}{Q}  \label{Walls-Millburn threshold}
\end{equation}%
where $v_{\max }$ is the threshold speed of the moving mirror in Figure 1,
or of the moving membrane in Figure 2, where $c$ is the vacuum speed of
light, and where $Q$ is the quality factor of the cavity for producing the
DCE. The maximum velocity amplitude of the moving membrane at threshold is
given by%
\begin{equation}
v_{\max }=\Omega \varepsilon _{\max }  \label{v_max = Omega epsilon_max}
\end{equation}%
where $\Omega $ is the angular frequency of the sinusoidal mechanical motion
of the moving mirror (i.e., of mirror M1 in Figure 1, or of the moving
membrane in Figure 2), and where $\varepsilon _{\max }\,\ $is the maximum
displacement in the sinusoidal motion of this mirror at threshold.

For superconducting radio frequency (\textquotedblleft SRF\textquotedblright
) cavities with $Q$ on the order of $10^{10}$ \cite{Walther}, we see the $%
v_{\max }$ will be on the order of centimeters per second, which is clearly
a non-relativistic velocity scale that is readily achievable under
laboratory conditions. Therefore although the generation of radiation is a $%
relativistic$ effect, the motion of the mirror that generates the DCE at its
threshold is \emph{non-relativistic} due to the high quality factors of SRF
cavities. One can understand the non-relativistic threshold condition (\ref%
{Walls-Millburn threshold}) as arising from a multiple-imaging effect, along
with its cumulative Doppler shifts, that occurs repetitively between the
moving mirror M1 and the fixed mirror M2 of the Fabry-Perot cavity in Figure
1 \cite{Chiao-images}.

Converting (\ref{Walls-Millburn threshold}) into an expression for the
kinetic energy in the motion of a mirror with a mass $m$, we find%
\begin{equation}
U_{\text{kin}}=\frac{1}{2}mv_{\max }^{2}\geq \frac{1}{2}\frac{mc^{2}}{Q^{2}}
\label{kinetic energy in M1}
\end{equation}%
If the mirror M1 in Figure 1 were to be driven on its left side by radiation
pressure from a pump wave stored inside a separate, high-$Q$ pump cavity on
the left side of M1 (not shown in Figure 1, but shown in Figure 2), then by
energy conservation, we find that the required amount of pump power that
needs to be injected into the pump cavity for the DCE at threshold, would be%
\begin{equation}
P_{\text{thres}}\geq \frac{U_{\text{kin}}}{\tau _{p}}=\omega _{p}\frac{U_{%
\text{kin}}}{Q_{p}}  \label{power in terms of ring-down time}
\end{equation}%
where $\tau _{p}$ is the \textquotedblleft cavity ring-down
time\textquotedblright\ for the energy stored inside the pump cavity. The
last equality follows from the fact that the pump-cavity quality factor $%
Q_{p}$\ is related to the pump cavity ring-down time $\tau _{p}$ by $%
Q_{p}=\omega _{p}\tau _{p}$ where $\omega _{p}$ is the angular frequency of
the pump wave

For the \textquotedblleft triple-cavity\textquotedblright\ paramp pictured
in Figure 2 whose membrane (red) is being pumped from the left by a
radiation pressure force, the frequency of the mechanical motion of this
moving membrane will be at the \emph{second harmonic} $2\omega _{p}$ of the
pump frequency. The meaning of the equality in (\ref{power in terms of
ring-down time}) is that in steady-state equilibrium, the amount of pump
power being injected into the \textquotedblleft single\textquotedblright\
pump cavity through the left porthole of this cavity, must be balanced by
the amount of mechanical power leaking away from the system due to the fact
that pump waves which are driving the motion of the membrane, will also be
escaping from the \textquotedblleft single\textquotedblright\ pump cavity
through the same porthole, or will be lost into heat.

Now if we set $Q_{p}=Q$ (i.e., that the pump and the DCE cavities to the
left and to the right of the moving membrane in Figure 2, will have
comparable $Q$ values), then it follows from (\ref{kinetic energy in M1})
and (\ref{power in terms of ring-down time}) that the injected pump power
for achieving threshold for the DCE should be%
\begin{equation}
P_{\text{thres}}\geq \frac{1}{2}\frac{m\omega _{p}c^{2}}{Q^{3}}
\label{power in terms of Q}
\end{equation}%
where $m$ is the mass of the moving mirror, $\omega _{p}$ is the pump
frequency, and $Q$ is the quality factor of cavities. Note that this DCE
threshold power scales inversely as the \emph{cube} of the $Q$ value of the
pump and the DCE cavities. This points out the importance of utilizing
cavities with the highest possible $Q$ values in order to be able to achieve
the DCE with reasonable pump powers. Therefore SRF cavities with $Q\sim
10^{10}$ \cite{Walther} would be good candidates for this purpose.

A more detailed derivation of the threshold power (\ref{power in terms of Q}%
) is given in \cite{Aharonov paper}, and yields the following result:%
\begin{equation}
P_{\text{thres}}\geq \frac{m\omega _{p}\omega _{s}\omega _{i}L^{2}}{%
4Q_{p}Q_{s}Q_{i}}  \label{threshold for 'triple' cavity}
\end{equation}%
where $m$ is the mass of the moving membrane, where $\omega _{p}$, $\omega
_{s}$, and $\omega _{i}$ are respectively, the pump, signal, and idler
frequencies of the \textquotedblleft triple\textquotedblright\ cavity
depicted in Figure 2 , where $L$ is the length of the \textquotedblleft
double\textquotedblright\ cavity in Figure 2, and where $Q_{p}$, $Q_{s}$,
and $Q_{i}$ are respectively, the pump, signal, and idler quality factors of
the three tandem SRF cavities that constitute the \textquotedblleft
triple\textquotedblright\ paramp cavity.

Numerically, if we assume that%
\begin{equation}
m=\text{ 3 milligrams}
\end{equation}%
\begin{equation}
\omega _{p}\approx \omega _{i}\approx \omega _{s}\approx 2\pi \times \text{%
10 GHz}
\end{equation}%
\begin{equation}
L\approx \lambda _{i}\approx \lambda _{s}\approx \text{3 cm}
\end{equation}%
\begin{equation}
Q_{p}\approx Q_{i}\approx Q_{s}\approx 10^{10}
\end{equation}%
then we conclude that for observing the DCE, and thus the generation of
gravitational microwave radiation, the threshold pump power at a frequency
of 10 GHz to be injected through the left hole of the \textquotedblleft
triple\textquotedblright\ paramp cavity of Figure 2, needs to be at least 
\begin{equation}
P_{\text{thres}}\approx \text{0.17 microwatts}
\end{equation}%
which is easily achievable experimentally.

\begin{figure}[]
\begin{center}
\includegraphics[width=1\textwidth]{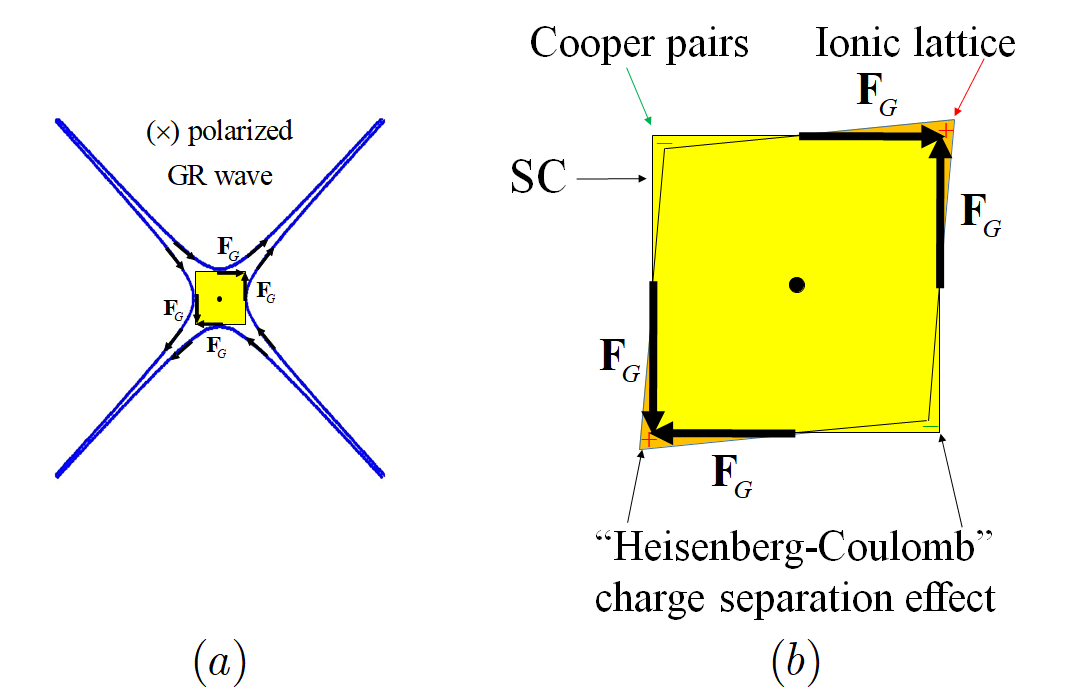} 
\end{center}
\caption{$(a)$ A quadrupolar pattern (blue) of an incident ($\times $)
polarized GR plane wave propagating into the page, impinges upon a square
piece of superconductor (yellow). Tidal \textquotedblleft
forces\textquotedblright\ $\mathbf{F}_{G}$\ acting upon the square due to
this wave\ exert a \emph{shear stress}. $(b)$ The \emph{strain} of the ionic
lattice of the square superconductor (SC) due to this \emph{stress} causes a
slight extrusion of positive ($+$) charge into corners of a rhombus (orange)
produced by the wave, but the Cooper pairs (yellow) of the SC will not
respond, since they are all Bose-condensed in a non-localizable,
zero-momentum eigenstate. There results an extrusion of negative ($-$)
charge into corners of an undistorted square (yellow), in a
\textquotedblleft Heisenberg-Coulomb\textquotedblright\ charge separation
effect \protect\cite{physica e}. The Coulomb attraction of the ($+$) and ($-$%
) charges opposes the tidal \textquotedblleft forces\textquotedblright\ $%
\mathbf{F}_{G}$, leading to a huge GR-EM coupling.}
\label{tidalforces}
\end{figure}

A crucial question now arises: How can we construct a high-Q cavity for
gravitational radiation, when we know that all known ordinary (that is,
non-astrophysical) materials, are essentially completely transparent to this
kind of radiation? To answer this question, consider a $\left( \times
\right) $ polarized GR plane wave incident upon a square piece of SC (yellow
square), as pictured in Figure 3$\left( a\right) $. The strain fields $%
h_{\mu \nu }$ of the incident GR wave will interact with the stress-energy
tensor $T^{\mu \nu }$\ of the SC via the interaction Hamiltonian density 
\cite{Dyson}%
\begin{equation}
H_{\text{interaction}}^{\prime }=\frac{1}{2}h_{\mu \nu }T^{\mu \nu }
\label{Dyson's interaction Hamiltonian}
\end{equation}

In particular, the instantaneous spatial components of the
transverse-traceless metric deviation tensor $h_{ij}^{\left( \times \right)
} $ for a $\left( \times \right) $ polarized plane wave, described in
Cartesian $\left( x,y\right) $ coordinates in a plane $z=$ constant
perpendicular to the $+z$ propagation direction of the wave, are given by
the following 2$\times $2 matrix \cite{MTW}:%
\begin{equation}
\left( h_{ij}^{\left( \times \right) }\right) =\left( 
\begin{array}{cc}
h_{xx} & h_{xy} \\ 
h_{yx} & h_{yy}%
\end{array}%
\right) =h_{0}\left( z-ct\right) \left( 
\begin{array}{cc}
+\frac{1}{2}\left( x^{2}-y^{2}\right) & xy \\ 
xy & -\frac{1}{2}\left( x^{2}-y^{2}\right)%
\end{array}%
\right)  \label{Ansatz for 2D plane wave solution}
\end{equation}%
where $\left( i,j\right) =\left( x,y\right) $ and where $h_{0}\left(
z-ct\right) $ is the dimensionless strain of space due to the passage of the
plane wave. A snapshot of the tidal \textquotedblleft
force\textquotedblright\ fields that are produced by the metric deviation
tensor $h_{ij}^{\left( \times \right) }\left( x,y,z,t\right) $ in (\ref%
{Ansatz for 2D plane wave solution}) is represented by the hyperbolae (blue
curves) in Figure 3$\left( a\right) $. One can easily verify by direct
substitution that the Ansatz given in (\ref{Ansatz for 2D plane wave
solution}) is a transverse-traceless vacuum solution to the wave equation
that follows from linearized general relativity, viz.,%
\begin{equation}
\nabla _{\bot }^{2}h_{ij}^{\left( \times \right) }+\frac{\partial
^{2}h_{ij}^{\left( \times \right) }}{\partial z^{2}}-\frac{1}{c^{2}}\frac{%
\partial ^{2}h_{ij}^{\left( \times \right) }}{\partial t^{2}}=0
\end{equation}%
where $\nabla _{\bot }^{2}$ is the transverse Laplacian in a Cartesian $%
\left( x,y\right) $ coordinate system, where $+z$ is the direction of
propagation of the plane wave $h_{ij}^{\left( \times \right) }\left(
x,y,z,t\right) $ (\ref{Ansatz for 2D plane wave solution}) into the page
that is depicted in Figure 3$\left( a\right) $, and where $c$ is the speed
of light.

The highly unusual $quantum$ response of the SC square (yellow) to this
wave, which we called the \textquotedblleft
Heisenberg-Coulomb\textquotedblright\ effect in \cite{physica e}, is
illustrated in Figure 3$\left( b\right) $. Quantum mechanics on a
macroscopic length scale inside the SC takes effect below the SC transition
temperature, due to the fact that Cooper pairs are bosons that will all
undergo Bose-Einstein condensation into the lowest possible energy state of
the system, namely the unique ground state in which all the bosons occupy
the same, single-particle \textit{zero-momentum eigenstate}, relative to the
center of mass of the SC (which is represented by the large black dot at the
center of the yellow square in Figure 3$\left( b\right) $). Because their
momenta will all be exactly known in the zero-momentum eigenstate (their
momenta will all certainly be $exactly$ zero), it follows from the
Heisenberg uncertainty relations for mometum and position, that the
locations of the Cooper pairs inside the SC square will be completely
uncertain. Thus the Cooper pairs are all $completely$ \emph{non-localizable}
within the SC square in Figure 3.

It therefore follows from Heisenberg's uncertainty principle that the Cooper
pairs cannot respond at all to the passage of the gravitational plane wave,
in contrast to the response to this wave of the ions inside the ionic
lattice, which are all $completely$ $localizable$, since they will be
located at the lattice sites of the ionic lattice inside the SC material.
Since the microwave frequencies of the incident gravitational plane wave in
Figure 3 are typically orders of magnitude higher that the typical acoustical
frequencies of the ionic lattice, it follows that the ions will move
essentially as freely falling masses along the geodesics of general
relativity, in their response to the passage of the plane wave. By contrast,
the Cooper pairs are \emph{completely nonlocalizable} due to the uncertainty
principle, and therefore it is forbidden in quantum mechanics for them to
follow $any$ classical trajectory, including the geodesics of general
relativity. This difference in response of the Cooper pairs and lattice ions has been demonstrated quantitatively in \cite{FQMT15}\cite{GRG essay}.

One can arrive at this same conclusion from another point of view. The
quantum adiabatic theorem tells us that for any SC sample, the BCS ground
state, which is separated from all possible excited states by the BCS energy
gap $E_{\text{BCS}}$, cannot respond to any slowly-varying external
perturbation whose characteristic frequency lies well below the BCS gap
frequency of $f_{\text{BCS}}=E_{\text{BCS}}/2\pi \hbar $. For the case of
niobium, $E_{\text{BCS}}$\ is around 3 meV, corresponding to a BCS gap
frequency $f_{\text{BCS}}\approx $ 730 GHz. Therefore any perturbations
arising from an incident GR wave whose typical frequency lies in the
microwave range of around 10 GHz, which is much less than 730 GHz, cannot
cause any transitions out of the BCS ground state. Hence the Cooper pairs
inside the SC (niobium) square of Figure 3 will remain rigidly in the BCS
ground state, and cannot respond to the incident GR microwaves at 10 GHz
that we are using in our experiments.

\begin{figure}[]
\begin{center}
\includegraphics[width=0.5\textwidth]{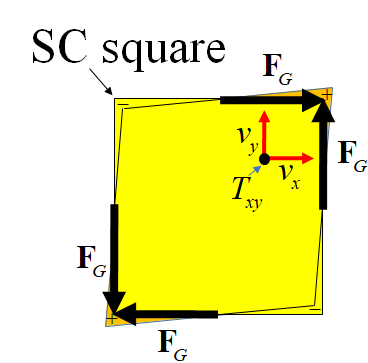} 
\end{center}
\caption{Gravitational Meissner-like effect inside a SC square (yellow) subjected
to tidal ``forces'' $\mathbf{F}_{G}$ from the $\left( \times \right) $ polarized
gravitational plane wave depicted in Figure 3. The stress-energy tensor $%
T_{xy}$ at a point along the main diagonal of the rhombus (orange) produced in response to the tidal \textquotedblleft
forces\textquotedblright\ $\mathbf{F}_{G}$, is a tensor product of the two
supercurrent vector components $v_{x}$ and $v_{y}$, both of which decay into
the interior on the scale of the London penetration depth.}
\label{grmeissner}
\end{figure}

However, the ions of the ionic lattice of the SC will undergo free fall in
response to the Newtonian tidal \textquotedblleft forces\textquotedblright\ $%
\mathbf{F}_{G}$ (i.e., the blue hyperbolae in Figure 3$\left( a\right) $).
Thus the ionic lattice will undergo a shear strain that distorts the SC
square (yellow) into a rhombus (orange), as shown in Figures 3 and 4. This
rhomboidal distortion leads to an extrusion of positive ionic charges into
the acute corners of the rhombus (orange corners labeled by $\left( +\right) 
$ signs). The overall charge of the SC, however, must remain neutral. Hence
the corners of the original square (yellow) (labeled by $\left( -\right) $
signs) adjacent to obtuse corners of the rhombus must have compensating
extrusions of negative charges arising from the negative charges of the
Cooper pairs that remain in these corners during the rhomboidal distortion
of the ionic lattice, because of the fact that these pairs must remain
adiabatically in their zero-momentum ground state everywhere.

There results a \textquotedblleft charge-separation
effect\textquotedblright\ (or \textquotedblleft Heisenberg-Coulomb
effect\textquotedblright ; see below)\ \cite{physica e}\cite{FQMT15}\cite{quach}, in which positive charges appear near the acute corners of the
rhombus of Figures 3 and 4, but negative charges appear near the obtuse
corners of this rhombus. This leads to a huge Coulomb force of attraction
between the separated positive and negative charges that strongly opposes
the Newtonian tidal \textquotedblleft forces\textquotedblright\ $\mathbf{F}%
_{G}$ of the incoming gravitational wave that produced this charge
separation in the first place. There arises an enormously stiff effective
Hooke's law, i.e., a strong restoring force inside the SC material, in that
there will arise an enormous Coulomb-strength back-action that strongly
resists the tidal action of the incoming gravitational plane wave, so much
so that this wave is $expelled$, and therefore $reflected$, from the SC
square, in what we shall call a \textquotedblleft gravitational Meissner
effect.\textquotedblright\ Since this effect results from a combination of
the Heisenberg uncertainty principle with the huge Coulomb force of
attraction that arises from the resulting separation of the ions from the
Cooper pairs, we have dubbed this the \textquotedblleft Heisenberg-Coulomb\
effect.\textquotedblright\ Therefore this effect differs from the usual
\textquotedblleft charge-separation effect\textquotedblright\ that occurs in
electrically polarized dielectrics in response to the application of an
electric field, because, firstly, it is a response to the $tensor$ $h_{ij}$
field of gravitational radiation, and not to the $vector$ electric field of
electromagnetism, and, secondly, this response is purely quantum mechanical
in nature, and possesses no classical explanation.

According to \cite{physica e}\cite{quach}\cite{Vector coupling}, the
strength of the \textquotedblleft Heisenberg-Coulomb\textquotedblright\
effect is characterized by the ratio of the strength of the Coulomb
electrical force between two electrons to the strength of their Newtonian
gravitational force%
\begin{equation}
\left\vert \frac{F_{\text{Coulomb}}}{F_{\text{Newton}}}\right\vert =\frac{%
e^{2}}{4\pi \varepsilon _{0}r^{2}\left( Gm_{e}^{2}/r^{2}\right) }=\frac{e^{2}%
}{4\pi \varepsilon _{0}Gm_{e}^{2}}\approx 4.2\times 10^{42}  \label{K_eff}
\end{equation}%
where $e$ is the electron charge, $\varepsilon _{0\text{ }}$is the
electrical permittivity of free space, $G$ is Newton's constant, and $m_{e}$
is the mass of the electron (note that the Coulomb and Newtonian forces both
obey inverse-law force laws, so that this result is independent of distance $%
r$ between the two electrons). The ratio given by (\ref{K_eff}) is obviously
a $huge$ dimensionless number.

One surprising consequence of the enormous number (\ref{K_eff}) predicted in 
\cite{physica e} is that it leads to such an enormous enhancement of the
reflection process from the SC square, that the SC behaves like a material
with hard-wall boundary conditions with respect to the incident
gravitational wave, and thus behaves like a highly reflective mirror. But
for such a hard-wall reflection to occur, it is necessary that the incident
gravitational wave would somehow generate sufficiently strong mass currents
on the surface of the mirror, such that these currents would then re-radiate
both a plane wave in the forwards direction that would cancel out the
incident wave, and would simultaneously re-radiate a plane wave in the
backwards direction that is 180 degrees out of phase with respect to the
incident wave, in order to create the totally reflected wave.

Due to the enormity of the \textquotedblleft Heisenberg-Coulomb
effect\textquotedblright\ predicted in (\ref{K_eff}), there could indeed
arise such enormous quantum-mechanical mass supercurrents, which are induced
by the extremely tiny strains of space associated with the incident
gravitational plane wave, so that even the tininess of Einstein's coupling
constant $8\pi G/c^{4}$ that couples sources to fields in Einstein's field
equations, might somehow be overcome during reflection. But how could one
possibly reconcile this with the Einstein's field equations without somehow
modifying its extremely tiny $8\pi G/c^{4}$\ coupling constant?

There already exists a hint as to how to handle this situation in
magnetostatics, in which the field equation in the vacuum is given by
Ampere's law%
\begin{equation}
\nabla ^{2}\mathbf{A}=-\mu _{0}\mathbf{j}
\label{magnetostatics in absence of magnetic materials}
\end{equation}%
where $\mathbf{A}$ is the vector potential from which the magnetic field is
derived, $\mu _{0}$ is the magnetic permeability of free space (i.e., the
vacuum without any medium), and $\mathbf{j}$ is the total current density,
which is the source of this field equation.

However, suppose that there exists a magnetic medium with a relative
magnetic permeability $\mu _{r}$, such as some ferromagnetic material that
fills all of space. It is a well known empirical fact that the insertion of
a high-permeability ferromagnetic material, such as iron, into the interior
of an electromagnet will greatly enhance the strength of the magnetic field
generated by this electromagnet. This empirical fact provides ample
justification for a modification of the field equation (\ref{magnetostatics
in absence of magnetic materials}), in which one inserts a prefactor $\mu
_{r}$ in front of the source-to-field coupling constant $\mu _{0}$, so that
this modified field equation now reads%
\begin{equation}
\nabla ^{2}\mathbf{A}=-\mu _{r}\mu _{0}\mathbf{j}
\label{magnetostatics in the presence of magnetic materials}
\end{equation}%
Thus in the presence of a homogeneous and isotropic magnetic medium, there
exists a dimensionless number $\mu _{r}$ (i.e., the \textquotedblleft
relative permeability\textquotedblright\ of the medium), which has a sign
and a magnitude that must be determined by experiment, as the prefactor of
the source term in the field equation (\ref{magnetostatics in the presence
of magnetic materials}).

Now for most \textquotedblleft normal\textquotedblright\ materials, it turns
out that the magnitude of $\mu _{r}$ is very close to unity. Both signs of
the permeability for magnetic materials (i.e., diamagnetic and paramagnetic
signs) exist in nature, but all of these permeabilities are quantum
mechanical, and not classical, in origin \cite{bohr}. In both cases of
diamagnetism and paramagnetism, quantum mechanical currents are required to
explain the phenomena. Moreover, in the case of ferromagnetic materials, $%
\left\vert \mu _{r}\right\vert $ has been observed to have very large
values, such as $10^{6}$ in iron-nickel alloys.

Note that one must carefully distinguish here between the \textquotedblleft
relative pemeability\textquotedblright\ and the \textquotedblleft relative
permittivity\textquotedblright\ of material media, because the magnetic
response of a given material is fundamentally different from its electric
response, since the magnetic field is fundamentally different in nature from
the electric field. Likewise, the question now arises: Does one need to make
similar distinctions in the case of the different possible responses of
various kinds of material media to the different kinds of gravitational
fields in general relativity?

We argue here that the answer to this question is yes. One reason for this
affirmative answer is that we know that in general relativity, there exists
the Lense-Thirring field, which is a \emph{gravito-magnetic} field, which is
fundamentally different in nature from the usual Newtonian gravitational
field, which is a \emph{gravito-electric} field. However, in addition to
these two kinds of fields, there exists in general relativity a third,
fundamentally different kind of field, namely, the transverse-traceless $%
h_{ij}$ \emph{gravito-tensor} field associated with gravitational radiation,
which has no analog in electromagnetism. In general relativity, we know that
the different components of the stress-energy tensor $T_{\mu \nu }$\ can be
sources for three different possible kinds of gravitational fields, and thus
in principle can lead to three different possible kinds of responses of
different material media to gravitational fields, namely, a scalar, a
vector, and a tensor response, which correspond to the components $T_{00}$, $%
T_{0i}$, and $T_{ij}$ of the tensor $T_{\mu \nu }$, respectively.

The modification of Ampere's law (\ref{magnetostatics in the presence of
magnetic materials}) in order to allow for the different possible responses
of homogeneous and isotropic magnetic media due to an applied magnetic $%
\mathbf{H}$ field arising from a solenoid, for example, justifies a similar
modification of Einstein's field equations, after they have been reduced to
a linearized wave equation for $h_{ij}$, in order to allow for the
possibility of different responses of homogeneous and isotropic material
media to a gravitational wave. In particular, there could arise enormous
quantum-mechanical mass supercurrents induced in a superconductor due to
even a tiny applied $T_{ij}$ stress field arising from the incident $\left(
\times \right) $ polarized plane wave depicted in Figures 3 and 4, which, in
light of the above \textquotedblleft Heisenberg-Coulomb\textquotedblright\
effect, would lead to internal electric fields inside the superconductor
that would drive these enormous supercurrents.

Before modification, the wave equation for gravitational waves is \cite%
{balbus}%
\begin{equation}
\nabla ^{2}h_{ij}-\frac{1}{c^{2}}\frac{\partial ^{2}h_{ij}}{\partial t^{2}}%
=-2\kappa _{0}T_{ij}  \label{wave equation in free space}
\end{equation}%
where the extremely tiny dimensionful constant 
\begin{equation}
\kappa _{0}=\frac{8\pi G}{c^{4}}  \label{Einstein's coupling constant}
\end{equation}%
is Einstein's coupling constant for the vacuum in the
absence of any medium. The dimensionful constant $\kappa _{0}$ is analogous
to the dimensionful constant $\mu _{0}$, the magnetic permeability of free
space (i.e., for the vacuum in the absence of any magnetic medium) in
Ampere's law (\ref{magnetostatics in absence of magnetic materials}).

After making the proposed modification, in which one inserts a prefactor $%
\kappa _{r}$ in front of the source-to-field coupling constant $\kappa _{0}$%
, the wave equation for gravitational waves now reads as follows:%
\begin{equation}
\nabla ^{2}h_{ij}-\frac{1}{c^{2}}\frac{\partial ^{2}h_{ij}}{\partial t^{2}}%
=-2\kappa _{r}\kappa _{0}T_{ij}  \label{wave equation in a medium}
\end{equation}%
where the dimensionless number $\kappa _{r}$ on
the right hand side of this wave equation \cite{not effective G}\cite%
{Abraham-Minkowski controversy}, is an empirically determined constant that
we shall call the \textquotedblleft relative gravitational
permeativity\textquotedblright\ \cite{permeativity}, in analogy with $\mu
_{r}$, the \textquotedblleft relative magnetic
permeability,\textquotedblright\ that was introduced into Ampere's law (\ref%
{magnetostatics in the presence of magnetic materials}). Although the
typical sizes of the relative permeability observed in ferromagnetic media $%
\left\vert \mu _{r}\right\vert \sim 10^{6}$ may not be as large as the
typical sizes of the relative gravitational permeativity $\left\vert \kappa
_{r}\right\vert $ that may eventually be observed in future experiments in
superconducting media, both the sign and the magnitude of $\kappa _{r}$ must
ultimately be arrived at $empirically$; they cannot be ruled out on any 
\textit{a priori} basis \cite{Press Szekeres and FH}.

Now for most \textquotedblleft normal\textquotedblright\ materials, the
relative gravitational permeativity $\kappa _{r}$ will most likely be very
close to unity, so that these media will be essentially completely
transparent to gravitational waves. Note, however, that the wave equation (%
\ref{wave equation in a medium}) is still $linear$, even after the inclusion
of the prefactor $\kappa _{r}$. This $linearity$ leads to the applicability
of the superposition principle for the solutions of this wave equation, and
also permits the resulting classical waves to be quantized using the
canonical quantization procedure outlined above.

For a superconductor, however, $\kappa _{r}$ may turn out to be huge.
Although an estimate based on an incorrect $vector$ coupling theory yields $%
\left\vert \kappa _{r}\right\vert \sim 10^{42}$ (as in (\ref{K_eff}) based
on \cite{physica e}\cite{Vector coupling}), both the sign and the magnitude
of this empirical constant must ultimately be determined by measurements,
such as via the Fresnel reflection coefficient $\left\vert \rho \left(
\omega \right) \right\vert ^{2}$\ off of the surface of a square plate,
which is given by%
\begin{equation}
\left\vert \rho \left( \omega \right) \right\vert ^{2}=\left\vert \frac{%
n\left( \omega \right) -1}{n\left( \omega \right) +1}\right\vert ^{2}
\label{Fresnel reflection coeff}
\end{equation}%
where $n\left( \omega \right) $ is given by a plasma-like formula for the
index of refraction, as shown in Appendix B. This measurement of $\left\vert
\rho \left( \omega \right) \right\vert ^{2}$, however, has never been
performed, since there exist at the present time no laboratory sources for
gravitational waves.

But perhaps the strongest reason for introducing the \textquotedblleft
relative gravitational permeativity\textquotedblright\ $\kappa _{r}$ into
the wave equation (\ref{wave equation in a medium}), would be the existence
of a \textquotedblleft gravitational Meissner-like effect.\textquotedblright\ To
this end, let us consider evaluating the stress-energy tensor $T_{xy}$
evaluated at a point along the major diagonal of the rhombus sketched in
Figures 3 and 4. Since any second-rank tensor can be written as a tensor
product of two vectors, one can always express $T_{xy}$ as the direct
product 
\begin{equation}
T_{xy}\propto v_{x}v_{y}  \label{T_xy propto v_x v_y}
\end{equation}%
where $v_{x}$ and $v_{y}$ are the $x$ and $y$ components of some vector
field inside the SC. But the only physically relevant vector field in the
problem at hand is the quantum-mechanical supercurrent velocity vector field
that is induced by the incident $\left( \times \right) $-polarized
gravitational plane wave sketched in Figure 3$\left( a\right) $.

Now the supercurrent velocity field $\mathbf{v}$ is directly proportional to
the supercurrent density $\mathbf{j}$, which in turn is directly
proportional to vector potential $\mathbf{A}$ via London's constitutive
relationship. This leads to the following proportionalities:%
\begin{equation}
\mathbf{A}\propto \mathbf{j}\propto \mathbf{v}  \label{A propto j propto v}
\end{equation}%
But Ampere's law leads to the following equalities:%
\begin{equation}
\nabla \times \mathbf{B}=\nabla \times \nabla \times \mathbf{A}=\mu _{0}%
\mathbf{j}  \label{Ampere's law}
\end{equation}%
Using the vector identity%
\begin{equation}
\nabla \times \nabla \times \mathbf{A=\nabla }\left( \nabla \cdot \mathbf{A}%
\right) -\nabla ^{2}\mathbf{A}
\end{equation}%
and using the London gauge $\nabla \cdot \mathbf{A}=0$, one then arrives at
London's equation, i.e., the following Yukawa-like equation with an
empirical constant $\kappa _{\text{L}}$:%
\begin{equation}
\nabla ^{2}\mathbf{A-}\kappa _{\text{L}}^{2}\mathbf{A=0}
\end{equation}%
which is a linear PDE. Using London's constitutive relations (\ref{A propto
j propto v}), we also arrive at the following Yukawa-like, linear PDE for
the supercurrent velocity field:%
\begin{equation}
\nabla ^{2}\mathbf{v-}\kappa _{\text{L}}^{2}\mathbf{v=0}
\end{equation}%
For the transverse supercurrent velocities flowing in the SC square
configurations of Figures 3 and 4, we find the following two PDE's:%
\begin{equation}
\frac{\partial ^{2}v_{x}}{\partial z^{2}}-\kappa _{\text{L}}^{2}v_{x}=0
\end{equation}%
\begin{equation*}
\frac{\partial ^{2}v_{y}}{\partial z^{2}}-\kappa _{\text{L}}^{2}v_{y}=0
\end{equation*}%
These equations possess the following exponentially decaying solutions:%
\begin{equation}
v_{x}\left( z\right) =v_{x}\left( 0\right) \exp \left( -\kappa _{\text{L}%
}z\right) =v_{x}\left( 0\right) \exp \left( -z/\lambda _{\text{L}}\right)
\label{v_x = ...}
\end{equation}%
\begin{equation}
v_{y}\left( z\right) =v_{y}\left( 0\right) \exp \left( -\kappa _{\text{L}%
}z\right) =v_{y}\left( 0\right) \exp \left( -z/\lambda _{\text{L}}\right)
\label{v_y = ...}
\end{equation}%
where the London penetration depth $\lambda _{\text{L}}$ is given by%
\begin{equation}
\lambda _{\text{L}}=\frac{1}{\kappa _{\text{L}}}
\label{London penetration depth}
\end{equation}%
For superconducting niobium, $\lambda _{\text{L}}$ is measured to be around 40 nm.

From the tensor product relationship (\ref{T_xy propto v_x v_y}) and from
the solutions for the supercurrent velocity field components (\ref{v_x = ...}%
) and (\ref{v_y = ...}), we conclude that the stress-energy tensor has the
following $z$ dependence%
\begin{equation}
T_{xy}\left( z\right) \propto v_{x}\left( z\right) v_{y}\left( z\right)
=\left( v_{x}\left( 0\right) \exp \left( -z/\lambda _{\text{L}}\right)
\right) \cdot \left( v_{y}\left( 0\right) \exp \left( -z/\lambda _{\text{L}%
}\right) \right)
\end{equation}%
Therefore it follows that the exponential decay solution for the
stress-energy tensor in the $z$ direction is given by%
\begin{equation}
T_{xy}\left( z\right) =T_{xy}\left( 0\right) \exp \left( -2z/\lambda _{\text{%
L}}\right) \propto \exp \left( -2z/\lambda _{\text{L}}\right)
\label{stress-energy tensor vs z}
\end{equation}%
so that $T_{xy}\left( z\right) $ decays $twice$ as fast as the supercurrent
velocity field into the depth of the SC. Therefore the exponential decay
length scale of $T_{xy}$, i.e., its gravitational penetration depth, is $%
half $ that of the electromagnetic London penetration depth (\ref{London
penetration depth}).

Now from the linearity of the wave equation (\ref{wave equation in a medium}%
) and from the solution (\ref{stress-energy tensor vs z}), we conclude that
the solution for the gravitational wave field $h_{xy}$ penetrating into the
SC square must also obey the proportionality relations%
\begin{equation}
h_{xy}\left( z\right) \propto T_{xy}\left( z\right) \propto \exp \left(
-2z/\lambda _{\text{L}}\right)
\end{equation}%
Therefore we conclude that the gravitational wave amplitude $h_{xy}$, like $%
T_{xy}$, decays $twice$ as fast as the supercurrent velocity field into the
depth of the SC. Hence the exponential decay length scale of gravitational
plane amplitude $h_{xy}\left( z\right) $ deep inside the SC is also $half$
that of the usual electromagnetic London penetration depth (\ref{London
penetration depth}), i.e., around 20 nm for the case of niobium. This is a
\textquotedblleft gravitational Meissner-like effect\textquotedblright\ that will
lead to the $expulsion$ of the incident gravitational plane wave from the
interior of the SC square in Figures 3 and 4, and therefore will lead to a
mirror-like, total $reflection$ of this wave.

\begin{figure}[]
\begin{center}
\includegraphics[width=0.9\textwidth]{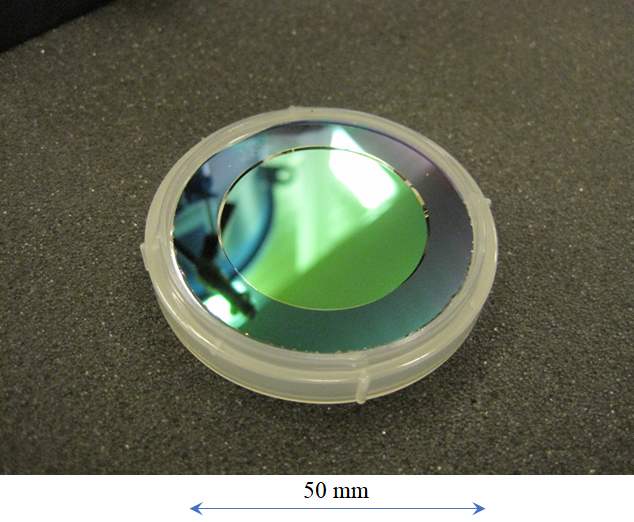} 
\end{center}
\caption{A flexible silicon nitride membrane (green; 500 nm thick) is
stretched over a circular window frame of an etched silicon wafer (gray; 50
mm diameter). A niobium coating (not shown) is sputtered onto the other side
of the membrane.}
\label{siliconnitridesample}
\end{figure}

\begin{figure}[]
\begin{center}
\includegraphics[width=0.9\textwidth]{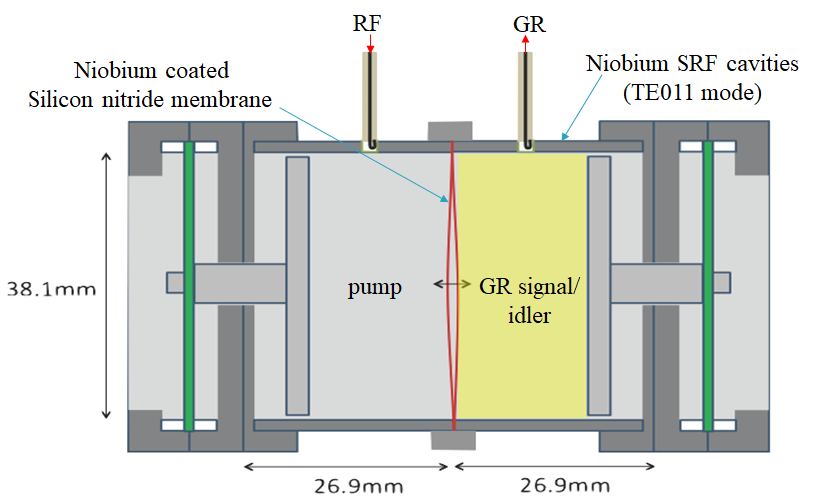} 
\end{center}
\caption{Degenerate parametric amplifier/oscillator design for generating
gravitational radiation (GR) in the signal/idler cavity (yellow) via the
mechanical motions (double-headed black arrow) of a silicon nitride membrane
coated with SC niobium (red) driven by microwaves in the pump cavity with
tuner (green).}
\label{degenerateparamp}
\end{figure}

\begin{figure}[]
\begin{center}
\includegraphics[width=0.80\textwidth]{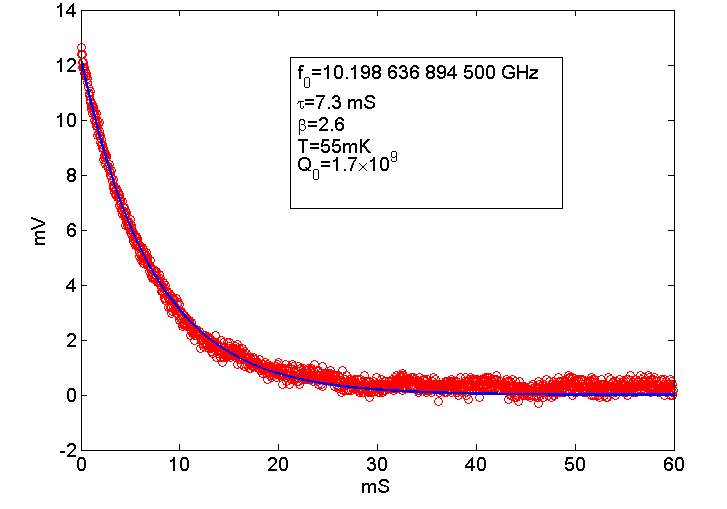} 
\end{center}
\caption{Exponential decay curve of a TEM microwave mode excitation of an
SRF stub cavity with a resonant frequency around 10 GHz. The exponential
ring-down time is 7.3 ms, implying a $Q$ of 1.7$\times 10^{9}$ at a
temperature of 55 mK.}
\label{ringdown}
\end{figure}

Now we present a progress report concerning our experiments towards
achieving the goal of observations of the DCE and of the laser-like
generation of gravitational waves. Figure 5 is a photograph of a silicon
nitride membrane sample, which is coated with niobium on its back side. \cite{NORCADA} This
membrane will be the active amplifying element in our paramps. We are
planning to place the sample shown in Figure 5 at the center of a degenerate
paramp as a vibrating membrane (red) driven by pump microwaves, as sketched
in Figure 6.

In this dual-SRF cavity setup, the pump injected into the left chamber is
identical in frequency to the signal and idler frequencies that will be
produced in the DCE in the right chamber (yellow in Figure 6) above a
certain threshold. Due to our prediction that the London penetration depth
for GR waves will be $half$ that for EM waves, the modal volume for the
right chamber at resonance will be slightly smaller for the case of GR wave
generation as compared to the case of EM wave generation. Hence there should
be a well-resolved difference in the position of the tuner (green) for EM
wave production relative to that for GR wave production inside the right
chamber. This difference will be a convenient signature that we can use to
distinguish between the two cases.

However, since the detection of GR waves will be difficult, we will first
try to $indirectly$ infer that these invisible waves are in fact being
generated by looking for a \textquotedblleft pump depletion
effect\textquotedblright\ in which there will arise a dip the reflected pump
signal from the left chamber at the threshold for GR wave generation. This
dip will arise from the \textquotedblleft missing energy\textquotedblright\
that will be escaping in the form of these invisible waves from the right
chamber. Thus we can infer from energy conservation that GR waves are in
fact being generated, although they will not be directly detectable.

In a follow-up experiment, we plan to make a copy of the degenerate paramp
apparatus pictured in Figure 6 as a \textquotedblleft
receiver,\textquotedblright\ and place it side-by-side with respect to the
\textquotedblleft transmitter,\textquotedblright\ in a Hertz-like
\textquotedblleft transmitter-receiver\textquotedblright\ configuration. The
ampilification of GR waves in the \textquotedblleft
receiver\textquotedblright\ paramp can serve as a low-noise preamp, i.e., a
first-stage amplifier, of a GR-wave detection system, whose final stage
could consist of a membrane-displacement measurement of a final-stage SC
membrane, whose displacement arises from the radiation pressure being
exerted on the membrane from the received GR waves.

In Figure 7, we show the progress that we have been making concerning the $Q$
problem. It turns out that gaps and other imperfections in the joints
between the cylindrical body of the SRF cavities and their endcaps can
degrade the $Q$ of the cavity by orders of magnitude. However, by
fabricating a seamless resonator using a coaxial stub cavity, one can evade
these kinds of degradations of the $Q$. Figure 7 is a plot of data from a
ring-down measurement of a SC niobium stub resonator that demonstrates that
we can achieve a $Q$ on the order of a billion at the typical temperature of
55 millikelvin that we have been using in our dilution refrigerators. If we
can achieve such a high $Q$ in the dual-SRF cavity sketched in Figure 6, we
will be well on our way towards demonstrating the DCE and, possibly, the
laser-like generation of gravitational waves.

It should be emphasized at this point that we are not trying to detect the
received GR waves by measuring the dimensionless $strain$ of space produced
by these waves, which would be exceedingly tiny, (see Appendix A), but
rather we shall try to detect the radiation pressure, and hence the received 
$power$, associated with these waves.\medskip

\bigskip \textbf{Appendix A: The strain of space produced by one milliwatt
of gravitational microwave power}

Since spacetime can be thought of as an extremely stiff medium, the question
naturally arises: How could one possibly produce any measurable amount of
strain of space, even if one were to succeed in a laser-like scheme for
generating gravitational (GR) waves? The short answer is this: One does not
need to be able to directly measure the $strain$ of space; one only needs to
be able to directly measure the $power$ in a laser-like beam of GR waves.
Nevertheless, it will be instructive to put in some numbers in order to
answer this question.

Suppose that one were able to generate one milliwatt of power in a
laser-like beam of a GR wave. The gravitational analog of the time-averaged
Poynting vector, which is the flux of energy, is given by \cite{Saulson}\cite%
{Townes MS}%
\begin{equation}
\left\langle S\right\rangle =\frac{\omega ^{2}c^{3}}{32\pi G}h_{\times }^{2}
\end{equation}%
where $h_{\times }$ is the strain of space for a $\left( \times \right) $
polarized plane wave. For one milliwatt of power in such a plane wave at 30
GHz, say, focused by a Newtonian SC telescope to a 1 cm$^{2}$ Gaussian beam
waist, one obtains a strain of space of%
\begin{equation}
h_{\times }\approx 0.8\times 10^{-28}
\end{equation}%
within the focal area. Such a tiny strain of space would be exceedingly
difficult to directly detect, even using advanced LIGO. However, it is
unnecessary to directly measure the $strain$ of space in order to detect the
GR wave, just as it is unnecessary to directly detect the \emph{optical} 
\emph{electric field amplitude} of a laser beam in order to detect the light
wave. Rather, one can directly measure the $power$ carried by the laser-like
GR beam, for example by measuring the back-conversion of one milliwatt of
the incident GR wave power into one milliwatt of EM wave power via a
measurement of the radiation pressure exerted by the received GR wave upon a
SC membrane in a time-reversed parametric process inside a replica of the
dual-SRF cavity of Figure 6. It would then be easy to detect one milliwatt
of the back-converted EM microwave power.

\medskip

\bigskip \textbf{Appendix B: Plasma-like gravitational-wave refractive
index of a superconductor}

The modified gravitational wave equation in a medium (such as a
superconductor (SC)) is%
\begin{equation}
\nabla ^{2}h_{ij}-\frac{1}{c^{2}}\frac{\partial ^{2}h_{ij}}{\partial t^{2}}%
=-2\kappa T_{ij}  \label{modified GR wave equation in a medium}
\end{equation}%
where we define%
\begin{equation}
\kappa \equiv \kappa _{r}\kappa _{0}
\end{equation}%
where $\kappa _{0}=8\pi G/c^{4}\ $is Einstein's coupling constant in vacuum,
and where $\kappa _{r}$ is the \textquotedblleft relative gravitational
permeativity\textquotedblright\ of the medium (to be determined by
experiment). We shall call $\kappa $ the \textquotedblleft gravitational
permeativity\textquotedblright\ of a SC medium, in analogy with the
\textquotedblleft magnetic permeability\textquotedblright\ of a magnetic
medium%
\begin{equation}
\mu \equiv \mu _{r}\mu _{0}  \label{mu = mu_r mu_0}
\end{equation}%
where $\mu _{r}$ is the relative magnetic permeability\ that appears as the
prefactor of the source term for Ampere's law in a medium%
\begin{equation}
\nabla ^{2}\mathbf{A}=-\mu \mathbf{j}  \label{ampere's law in a medium}
\end{equation}

Let us define the constitutive relation of a SC medium as follows:%
\begin{equation}
T_{ij}\equiv -\mu _{G}h_{ij}  \label{T_ij = ...}
\end{equation}%
where $\mu _{G}$\ is the \textquotedblleft gravitational shear modulus" of
the material to an applied $h_{ij}$ field.%
\ Substituting this constitutive relation into the modified wave equation in
a medium (\ref{modified GR wave equation in a medium}), we find%
\begin{equation}
\nabla ^{2}h_{ij}-\frac{1}{c^{2}}\frac{\partial ^{2}h_{ij}}{\partial t^{2}}%
=-2\kappa \mu _{G}h_{ij}  \label{wave equation with constitutive relation}
\end{equation}%
Upon substitution of the monochromatic plane wave Ansatz,%
\begin{equation}
h_{ij}\left( x,y,z,t\right) =A\exp \left( ikz-i\omega t\right)
\label{monochromatic plane wave ansatz}
\end{equation}%
into this equation, we obtain the implicit dispersion relation%
\begin{equation}
k^{2}-\frac{\omega ^{2}}{c^{2}}=-2\kappa \mu _{G}  \label{dispersion relation}
\end{equation}%
Let us now define the \textquotedblleft gravitational plasma
frequency\textquotedblright\ as%
\begin{equation}
\omega _{G}\equiv \sqrt{2\kappa c^{2}\mu _{G}}
\label{gravitational plasma frequency}
\end{equation}%
This agrees with \cite{FQMT15} since%
\begin{equation}
\kappa =\kappa _{r}\kappa _{0}  \label{G replaced by kappa_r G}
\end{equation}

Solving for $k\left( \omega \right) $ from (\ref{dispersion relation}), one
finds the explicit dispersion relation%
\begin{equation}
k\left( \omega \right) =\frac{\omega }{c}\sqrt{1-\frac{\omega _{G}^{2}}{%
\omega ^{2}}}  \label{k(w) = ...}
\end{equation}%
from which we see that the meaning of the plasma frequency is that%
\begin{equation}
k\left( \omega _{G}\right) =0  \label{k(w_P) = 0}
\end{equation}%
i.e., that the plasma frequency is a cutoff frequency below which a
gravitational wave cannot propagate inside the SC medium, because the
propagation wavenumber $k\left( \omega \right) $ becomes a pure imaginary
quantity.

Alternatively, let us introduce the index of refraction $n\left( \omega
\right) $ as follows:%
\begin{equation}
k\left( \omega \right) =\frac{n\left( \omega \right) \omega }{c}
\label{index of refraction}
\end{equation}%
Comparing this with (\ref{k(w) = ...}), we see that%
\begin{equation}
n\left( \omega \right) =\sqrt{1-\frac{\omega _{G}^{2}}{\omega ^{2}}}
\label{n(w) is plasma-like}
\end{equation}%
which is a plasma-like index of refraction. Note that for $\omega <\omega
_{G}$, the refractive index becomes a pure imaginary quantity, which implies
total reflection, just like the reflection from a plasma of an EM wave whose
frequency is below cutoff.

Thus the Fresnel reflection formula (\ref{Fresnel reflection coeff}) is \cite{Born&Wolf}%
\begin{equation}
\left\vert \rho \left( \omega \right) \right\vert ^{2}=\left\vert \frac{%
n\left( \omega \right) -1}{n\left( \omega \right) +1}\right\vert ^{2}
\label{corrected Fresnel formula}
\end{equation}%
where $n\left( \omega \right) $ is given by the plasma-like formula for the
index of refraction (\ref{n(w) is plasma-like}).\medskip

\textbf{Acknowledgments:} This work was supported in part by DARPA. We thank
Professors Douglas Singleton and Gerardo Mu\~{n}oz for their help on the
theory, and Jacob Pate for his help on the experiments. This paper, based on
a talk given on November 3, 2017 by RYC at the Aerospace Advanced Propulsion
Workshop, will appear in the Journal of the British Interplanetary
Society.


\begin{thebibliography}{99}
\bibitem{swing} Parametric amplification arising from the pumping motion of
the moving mirror is much like the result of the pumping action of a child
repeatedly standing and squatting on a swing, whose center of mass moves up
and down at $twice$ the natural frequency of the swing.
\bibitem{paramp kinematics} For the \textquotedblleft
triple-cavity\textquotedblright\ paramp depicted in Figure 2, the pump,
signal, and idler frequencies obey the relationship $2\omega _{p}=\omega
_{s}+\omega _{i}$, where the \textquotedblleft double\textquotedblright\
cavity possesses a doublet spectrum $\left( \omega _{s},\omega _{i}\right) $%
, whose splitting frequency is determined by the size of the iris in the
middle of this \textquotedblleft double\textquotedblright\ cavity. This
paramp is a phase-sensitive amplifier that obeys the phase relationships $%
2\phi _{p}=\phi _{s}+\phi _{i}\pm \pi /2$, where the phase $-\pi /2$
corresponds to maximum amplification, and $+\pi /2$ corresponds to maximum
deamplification \cite{Aharonov paper}.
\bibitem{Vector coupling} However, the enormous enhancement found in \cite%
{physica e}\ of the reflectivity of a SC mirror by huge dimensionless ratio $%
\left\vert F_{\text{Coulomb}}/F_{\text{Newton}}\right\vert \sim 10^{42}$ was
derived on the basis of the incorrect $vector$ interaction Hamiltonian of
the form $\mathbf{p\cdot A}$ between the Cooper pairs and $vector$
gravitational fields, and not on the basis of the correct $tensor$
interaction Hamiltonian of the form $h_{\mu \nu }T^{\mu \nu }$ between the
Cooper pairs and $tensor$ gravitational fields used in \cite{FQMT15}.
\bibitem{not effective G} Note that the introduction of the relative
gravitational permeativity $\kappa _{r}$ into the source of the wave
equation (\ref{wave equation in a medium}) affects solely the source of the
tensor $h_{ij}$ field associated with gravitational waves. This is not
equivalent to the introduction of an effective Newton's gravitational
constant $G_{eff}$ to replace Newton's constant $G$ in front of the source
term in the full Einstein's field equations so that they would now read as
follows:%
\begin{equation}
R_{\mu \nu }-\frac{1}{2}g_{\mu \nu }R=\frac{8\pi G_{eff}}{c^{4}}T_{\mu \nu }
\label{Einstein's field equation with G_eff}
\end{equation}%
since this replacement $G\rightarrow G_{eff}$\ would affect $all$ the
components of $T_{\mu \nu }$ equally. Such a replacement makes no
distinction at all between the scalar, vector, or tensor sources $T_{00}$, $%
T_{0i}$, or $T_{ij}$ of the three different kinds of gravitational fields.
However, the Newtonian gravitational field of the Earth, whose source is $%
T_{00}$, will not be affected by the introduction of the relative
gravitational permeativity $\kappa _{r}$ in (\ref{wave equation in a medium}%
). Hence there cannot be any \textquotedblleft
anti-gravity\textquotedblright\ arising from the presence of superconductors.
\end{thebibliography}
\end{document}